\documentclass[rmp,floatfix]{revtex4}

\usepackage{natbib}
\usepackage{times,setspace}
\usepackage{amsmath, graphicx, amssymb,tabls}
\usepackage{graphicx}
\usepackage{mathptmx}
\usepackage{amsmath,amssymb, amsthm}

\usepackage{psfrag}
\usepackage{color}

\newcommand{\beq}{\begin{eqnarray}}
\newcommand{\eeq}{\end{eqnarray}}
\newcommand{\beqn}{\begin{equation}}
\newcommand{\eeqn}{\end{equation}}
\newcommand{\bal}[1]{\begin{align} #1 \end{align}}

\newcommand{\expt}[1]{\langle #1 \rangle }

\newcommand{\ra}{\rightarrow}

\newcommand{\LL}{ {\mathcal{L}} }
\newcommand{\II}{\mathcal{I}}

\begin{document}

 \title{Sample genealogies and genetic variation in populations\\ of variable size}

\author{A. Eriksson$^{a,b}$, B. Mehlig$^{a}$, M.Rafajlovic$^{a}$, and S. Sagitov$^{c}$\\
$^a$\emph{\small Department of Physics, University of Gothenburg, SE-41296 Gothenburg, Sweden}\\
$^b$\emph{\small Department of Zoology, University of Cambridge, Cambridge, UK}\\
$^c$\emph{\small Mathematical Sciences, Chalmers University of Technology and University of Gothenburg, \\\hspace*{2mm}SE-41296 Gothenburg, Sweden}\\}

\begin{abstract}
We consider neutral evolution of a large population subject to chan\-ges
in its population size. For a population with a time-variable carrying capacity
we have  computed the distributions of the total branch lengths of its sample
genealogies. Within the coalescent approximation we have obtained a general expression -- Eq.~(\ref{eq:corr2}) --
for the moments of these distributions for an arbitrary smooth dependence of
the population size on time.
We investigate  how the
frequency of population-size variations alters  the distributions.
This allows us to discuss their influence on the distribution
 of the number of mutations, and on the population homozygosity in populations
with  variable size.
\end{abstract}
\maketitle

\newpage
\section{Introduction}
Models for gene genealogies of biological populations often assume a constant, time-independent
population size $N$. This is the case for the Wright-Fisher model \citep{Wright:1931,Fisher:1930}, for the Moran model
\citep{Moran:1958}, and for their representation in terms of the coalescent \citep{Kingman:1982}.
In real biological populations, by contrast, the population size changes
over time. Such fluctuations may be due to
catastrophic events (bottlenecks) and subsequent population expansions, or just reflect the randomness in the factors determining
the population dynamics. Many authors 
have argued that genetic variation in a population subject to size fluctuations
may nevertheless be described by the Wright-Fisher model, if one replaces the constant population size in this model by
an effective population size of the form
\begin{equation}
\label{eq:Neff}
N_{\rm eff} = \left(\lim_{T\rightarrow\infty}\frac{1}{T} \int_0^T\frac{{\rm d} t}{N(t)}\right)^{-1}
\end{equation}
(see, e.g., \citet{ewe82:neff} for a review of different measures of the effective population size, and \citet{Sjodin:2005,Wakeley:2009} for
recent extensions of this concept).
The harmonic average in Eq.~(\ref{eq:Neff}) is argued to capture the significant effect of catastrophic events on
 patterns of genetic variation in a population: if for example a population went through a 
a recent bottleneck, a large fraction of individuals in a given sample would originate
from few parents. This in turn would lead to 
significantly reduced genetic variation, parameterised by a small
value of $N_{\rm eff} $.

The concept of an effective population size has been frequently used in the literature, implicitly assuming
that the distribution of neutral mutations in a large population of fluctuating size is identical
to the distribution in a Wright-Fisher model with the corresponding constant effective population
size given by Eq.~(\ref{eq:Neff}). However, 
recently it has been shown that this is true only
under certain circumstances \citep{Nordborg:2003,Kaj:2003,JagersSagitov:2004}. 
It is argued by \citet{Sjodin:2005} that the concept of an effective
population size is appropriate when the time scale of fluctuations of $N(t)$ is either much smaller or much larger
than the typical time between coalescent events in the sample genealogy. In these limits it can be proven
that the distribution of the sample genealogies
is exactly given by that of the coalescent with a constant, effective
population size.

\begin{figure}[t]
\psfrag{y}[][t]{$\rho(T_n)$}
\psfrag{x}[t][]{$T_n$}
\psfrag{a}[][]{\textbf{a}}
\psfrag{b}[][]{\textbf{b}}
\psfrag{c}[][]{\textbf{c}}
\psfrag{d}[][]{\textbf{d}}
\psfrag{e}[][]{\textbf{e}}

\centerline{\includegraphics[width=8cm]{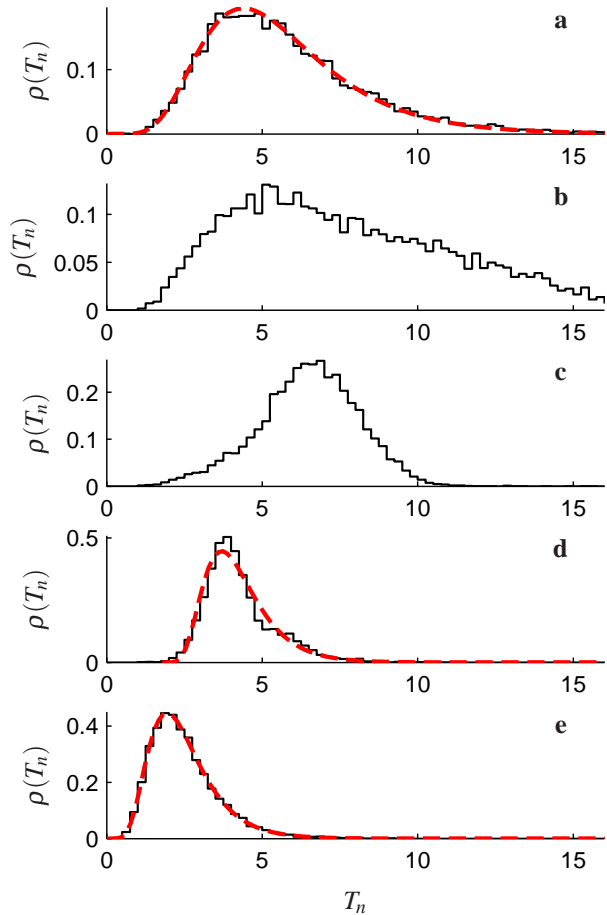}}
\caption{\label{fig:dist} 
Numerically computed distributions $\varrho(T_n)$ of 
total branch lengths $T_n$ in genealogies of samples of size $n=10$.
The model employed in the simulations is outlined
in Sec.~\ref{sec:sim}. It describes
a population subject to a time-varying carrying capacity. 
Panels {\bf a} to {\bf e} show $\varrho(T_n)$ for populations
with increasingly rapidly oscillating carrying capacity.
The dashed red line in {\bf a} shows a low-frequency approximation
to $\rho(T_n)$ obtained for a constant carrying capacity. 
The dashed red lines in {\bf d} and {\bf e} show the large-frequency approximations 
given by Eq.~(\ref{eq:large_w_approx}).
Further numerical and analytical results on the frequency dependence of
the moments of these distributions are shown in Fig. \ref{fig:stoch_moments}. 
Parameter values (see Sec. \ref{sec:sim} for details): $K_0 = 10,000$, $r = 1$, $\epsilon = 0.9$, and
{\bf a} $\nu K_0 = 0.001$,
{\bf b} $\nu K_0 = 0.1$,
{\bf c} $\nu K_0 = 0.316$,
{\bf d} $\nu K_0 = 1$, and
{\bf e} $\nu K_0 = 100$.
}
\end{figure}

More importantly, it follows from these results that, in populations with variable size, the coalescent with 
a constant effective population size
is not always a valid approximation for the sample genealogies.
Deviations between the predictions of the standard coalescent model and
empirical data are frequently observed, 
and there is a number of different statistical tests quantifying the corresponding discrepancies 
(see for example \citep{taj89:sta,FuLi:1993,ZengFuShiWu:2006}).
The analysis of such deviations is of crucial importance in understanding for example
human genetic history \citep{GarriganHammer:2006}.
But while there is a substantial amount of work numerically quantifying deviations, 
often in terms of a single number,
little is known about their qualitative  origins and their effect upon summary  statistics in the population in question.

\begin{figure}[t]
\centerline{\includegraphics[angle=0,width=10cm,clip]{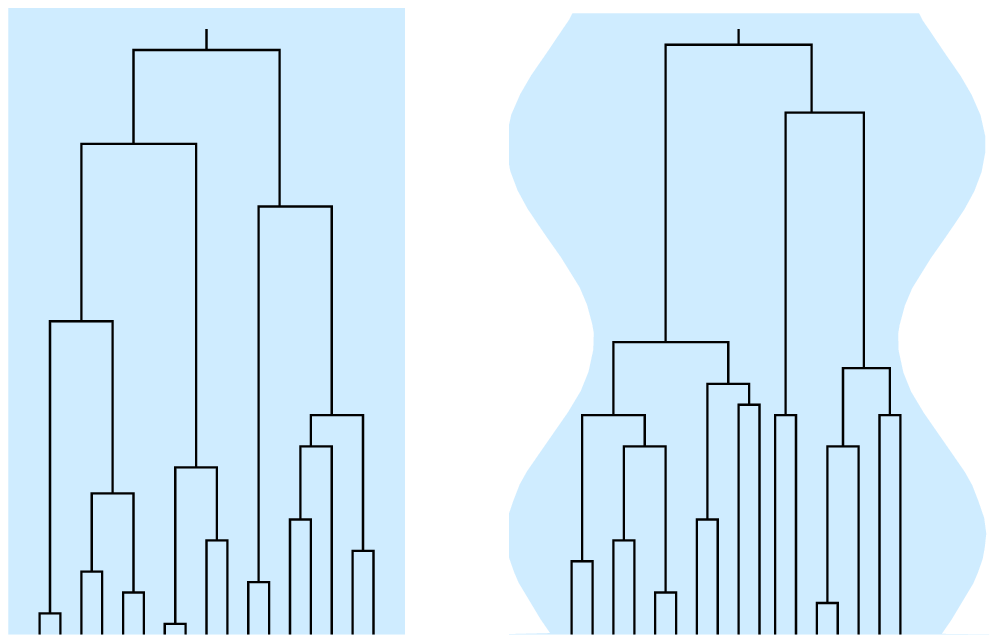}}
\caption{\label{fig:examples} Illustrates the effect of population-size
oscillations on the genealogy of a sample of size $n=17$ (schematic). Left: genealogy
described by Kingman's coalescent for a large population of constant size, 
illustrated by the light blue rectangle. Right: sinusoidally varying population size. Coalescence is accelerated in regions of
small population sizes, and vice versa. This significantly alters the tree and gives rise to changes in the distribution of 
the number of mutations, and of the population homozygosity.}
\end{figure}

The aim of this paper is to study the effect of population-size fluctuations on the patterns of genetic variation
for the case where the scale of the population-size fluctuations is comparable to the time
between coalescent events in the ancestral tree. As is well-known,
empirical measures of genetic variation can usually be computed from the total branch length
of the sample genealogy (the expected number of single-nucleotide polymorphisms, for example, is proportional
to the average total branch length). In the following we therefore analyse the distributions of 
the total branch lengths for sample genealogies in a population of fluctuating size.
An example is given in Fig. \ref{fig:dist} which shows numerically computed branch-length
distributions for a particular model population (described in Sec. \ref{sec:sim})
with a time-dependent carrying capacity.

As Fig. \ref{fig:dist} shows, the distributions depend in a complex manner on the form of the size changes.
We observe  that when the frequency of the population-size fluctuations 
is either very small or very large, the results are well
described by Kingman's coalescent with a constant (effective) population size.
Apart from these special limits, however,
the form of the distributions  appears to depend in a complicated manner upon the frequency
of the population-size variation.
The observed behaviour is caused by the 
fact that coalescence proceeds faster for smaller population sizes, and more slowly for larger population sizes, as 
illustrated in Fig.~\ref{fig:examples}. But the question is how to quantitatively account for
the changes displayed in Fig. \ref{fig:dist}.

We show in this paper that the results of the simulations shown in Fig. \ref{fig:dist}
are explained by a general expression -- Eq.~(\ref{eq:corr2}) --
for the moments of the distributions shown in Fig.~\ref{fig:dist}.
Our general result is obtained within the coalescent approximation valid in the limit of
large population size. But 
we find that in most cases, the coalescent approximation
works very well down to small population sizes (a few hundreds of individuals).
Our result enables us to understand and quantitatively describe
the frequency dependencies of the distributions shown in Fig.~\ref{fig:dist}. 
It makes  possible to determine for example
how the variance, skewness, and the kurtosis of these distributions depend upon the frequency
of demographic fluctuations. 
This in turn allows us to compute the population homozygosity and 
to characterise genetic variation in populations with size fluctuations.

The remainder of this paper is organised as follows.
In Sec.~\ref{sec:obs}
we review how empirical observables are related to the branch lengths
of the sample genealogies. Section~\ref{sec:rec} summarises
our analytical results for the moments of the total branch length.
In Sec.~\ref{sec:sim} we describe the model employed in the computer
simulations. The corresponding numerical results are compared to
the analytical predictions in Sec. \ref{sec:num}.
Finally,  in Sec.~\ref{sec:conclusions} we summarise how population-size
fluctuations influence the distribution of total branch lengths, discuss the implications for
patterns of genetic variation, and conclude with an outlook.

\section{Observables}
\label{sec:obs}
In this section we review how empirical observables are related to the branch lengths
of the sample genealogies.

Patterns of genetic variation reflect the gene genealogy
corresponding to a given sample. Within a neutral infinite-sites model,
mutations are assumed to occur randomly at a constant
rate $\mu$ on the genealogy. 
For a sample of size $n$, 
the  number $ S_n$ of 
single-nucleotide polymorphisms
conditioned on the total branch length $T_n$
of the sample genealogy has a Poisson distribution
with mean $\theta T_n/2$ (here $\theta$ is a scaled mutation parameter,
$\theta = 2 \mu N_0$ where $N_0$ is a suitable measure
of the population size):
\begin{equation}
\langle S_n \rangle = \frac{\theta}{2}\, \langle T_n\rangle\,.
\label{eq:theta}
\end{equation}
Similarly, moments of $S_n$ can be computed in terms of moments of $T_n$.
As is well known, the corresponding relations are most conveniently expressed in terms
of the function $F_n(q)$ from which the  moments can be computed
by repeated differentiation with respect to $q$:
\begin{equation}
\label{eq:genf}
F_n(q)
= \langle {\rm e}^{-q T_n}\rangle\,,\quad
\mbox{so that}\quad \langle T_n^k\rangle = (-1)^k
\frac{{\rm d}^k}{{\rm d}q^k} F_n(0)\,.
\end{equation}
Note that $F_n(\theta/2)$ is the probability of observing no mutations
in a sample of size $n$ (thus $F_2(\theta/2)$ is the population homozygosity).

The corresponding function for the moments of $S_n$ is found to be:
\begin{equation}
\label{eq:pSn}
 \langle {\rm e}^{-q S_n}\rangle
= F_n\big(\frac{\theta}{2}(1-{\rm e}^{-q})\big)\,.
\end{equation}
For a constant population size, this equation is equivalent
to Eq. (1.3a) in \citep{Watterson:1975}.

In short, the distribution of single-nucleotide polymorphisms is determined
by the function $F_n(q)$, or equivalently
by the moments $\langle T_n^k\rangle$.

Microsatellite loci by contrast are usually modeled in terms of a step-wise  mutation model \citep{Ota:1973} in which
a mutation corresponds to either the gain or, equally likely, the loss of a repeat unit. Provided 
that such steps (mutations) occur  according to a Poisson process, the distribution
of the  difference $j$ in the numbers of repeats between two randomly sampled
sequences is determined by the function $F_2$ \citep{Ota:1973,kimmel1996measures}:
\bal{
\label{eq:pj}
	p_j = \frac{1}{2\pi} \int_0^{2\pi} \!\!{\rm d}\omega\,\cos(\omega j) \,
       F_2\big(\theta(1 - \cos \omega)\big)\,.
}
In summary, the function $F_n$ (or equivalently the moments $\langle T^k_n\rangle$) allow
to compute the statistical fluctuations of the numbers
of single-nucleotide polymorphisms and of the number of steps in a step-wise
 mutation model.
In Sec.~\ref{sec:rec} we show how the moments and the function $F_n(q)$ may be determined
for a large neutral population subject to smooth population-size changes of otherwise arbitrary form.

\section{Coalescent approximation formulae for $F_n(q)$ and $\langle T_n^k\rangle$}
\label{sec:rec}
In this section we show how to calculate the function $F_n(q)$ and the moments
$\langle T_n^k\rangle$ within the coalescent approximation, for
a population with a smoothly varying size.

For $q=\theta/2$, 
the quantity $F_{n-1}(\theta/2)$
is just the probability that $n-1$  sequences sampled at the 
present time are identical. Thus in a population of constant size, 
$F_n(q)$ is given by
\begin{equation}
F_n(q)=\frac{\binom{n}{2}} {\binom{n}{2}+nq}  F_{n-1}(q)\,.
\end{equation}
This recursion has the well-known solution (with initial condition $F_1=1$)
\begin{equation}
\label{eq:Fn}
F_n(q) = \frac{\Gamma(n)\Gamma(1+2q)}{\Gamma(n+2q)}\,.
\end{equation}

The question is how to obtain a corresponding expression for the case of a changing
population size. 
We assume that in the limit of large population sizes, the changes
in population size are described by a smooth curve 
$x(t) = N(t)/N_0$, so that $0 < x(t) \leq 1$ and $x(0)=x_0$ at the present time, $t=0$.
As common in the coalescent approximation time is counted \lq backwards', that is
$t$ increases from the present ($t=0$) to the past.

\begin{figure}
\centerline{\includegraphics[angle=0,width=12cm,clip]{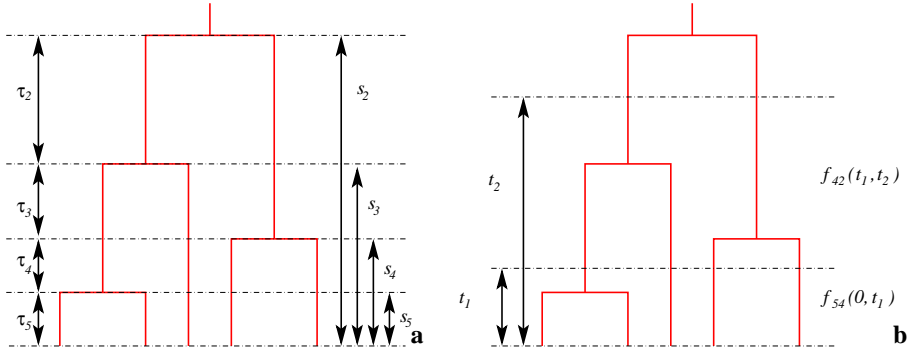}}
\caption{\label{fig:theta} 
{\bf a} Illustrates the definition of the variables $\tau_j$ and 
$s_j = \sum_{k=j}^n \tau_k$ used in the calculation of $F_n(q)$.  
In the example given, the sample size is $n=5$.
{\bf b} Illustrates the definition
of the times in Eq.~(\ref{eq:tij2}).
}
\end{figure}
Given a realisation of the curve $x(t)$, the function $F_n(q)$ can
be calculated as follows. 
The starting point is
the joint distribution of times $\tau_j$ 
(illustrated in Fig.~\ref{fig:theta}{\bf a}). 
As shown by \citet{GriTav:94} it can be written in terms the variables $s_j = \sum_{k=j}^n \tau_k$:
\begin{equation}
\label{eq:ft}
f(\tau_2,\ldots,\tau_n) = 
\prod_{j=2}^n b_j x(s_j)^{-1} {\rm e}^{-b_j[\Lambda(s_j)-\Lambda(s_{j+1})]}\,.
\end{equation}
 Here $b_j = j(j-1)/2$ and
$ \Lambda (t) =    \int_0^{t}{{\rm d}t'}{x(t')}^{-1}$
 is the \lq population-size intensity function' defined by \citet{GriTav:94}.
The distribution of the times $\tau_j$ during which the sample genealogy has $j$ lines 
depends upon the sample size $n$. This dependence is not made
explicit here, neither in Eq. (\ref{eq:ft}) nor in the following.
The corresponding joint density for the variables $s_j$ is simply   \citep{Tav:2004}
\begin{equation}
g(s_2,\ldots,s_n) = \prod_{j=2}^n b_j {x(s_j)}^{-1} {\rm e}^{-b_j[\Lambda(s_j)-\Lambda(s_{j+1})]}
\end{equation}
(for $0 < s_n < s_{n-1} < \cdots < s_2$, and $s_{n+1}=0$).

Now we make use of the fact that the total time is given by
\begin{equation}
T_n = s_n + s_{n-1} + \ldots + s_3 + 2 s_2\,.
\end{equation}
(see Fig. \ref{fig:theta}{\bf a}). The function $F_n(q)$ can therefore be written as
\begin{eqnarray}
F_n(q) &=& \int_{0 < s_n < s_{n-1} < \cdots < s_2} {\rm d}s_n \cdots {\rm d}s_2\, g(s_2,\ldots,s_n)\, {\rm e}^{-q(s_n + s_{n-1} + \ldots + s_3 + 2 s_2)}\,.
\end{eqnarray}
Expanding the multiple integrals one obtains
\begin{equation}
\label{eq:r2}
F_n(q) = b_n \int_0^\infty \frac{{\rm d}s_n}{x(s_n)} {\rm e}^{-[(n-1)\Lambda(s_n)+q s_n]} \cdots 
        b_2 \int_{s_3}^\infty \frac{{\rm d}s_2}{x(s_2)} {\rm e}^{-[\Lambda(s_2)+2q s_2]}\,.
\end{equation}
For small sample sizes $n$, Eq.~(\ref{eq:r2}) provides a convenient way
of computing the function $F_n(q)$ and the corresponding moments
$ \langle T_n^k\rangle$. For example, for $n=2$ one finds simply
\begin{eqnarray}
F_2(q)&=&1-2q\int_0^\infty dt \, e^{-\Lambda(t)-2q t}\,,\\
\langle T_2^k\rangle &=&  2^k\,k\int_0^\infty\!\!{\rm d}t\,\, t^{k-1} \, {\rm e}^{-\Lambda(t)}\,.
\label{eq:T2k}
\end{eqnarray}
This makes it possible to compute the population homozygosity in large populations
with arbitrary size variations, as well as the distribution
of steps in a step-wise mutation model, according to Eq.~(\ref{eq:pj}).

For large values of $n$, by contrast,
the large number of nested integrals in (\ref{eq:r2})
becomes increasingly difficult to evaluate. In this limit,
however, the distribution is 
conveniently characterised in terms of its cumulants which can
be expressed in terms of the moments $\langle T_n^k\rangle$.

In the remainder of this section we show how to calculate
the moments for arbitrary sample sizes $n$. According to
Eq.~(\ref{eq:genf}), these moments are obtained by repeated differentiation of 
Eq.~(\ref{eq:r2}). 
However, in the following we describe a more elegant approach making
use of a result obtained by \citet{Tav:84}.
As Fig.~\ref{fig:theta}{\bf a} shows, an alternative expression for the total time is simply 
$T_n=\sum_{m=2}^n m \tau_m$. The $k$-th moment of the distribution of $T_n$ is therefore 
\begin{equation}\label{eq:Tnk2}
\langle T_n^k\rangle = 
\sum_{\stackrel{\nu_2,\nu_3,\ldots,\nu_n}{\nu_2+\nu_3+\cdots+\nu_n=k}}
\binom{k}{\nu_2,\nu_3,\ldots,\nu_n}  
n^{\nu_n}\cdots 2^{\nu_2}\,\langle \tau_n^{\nu_n}
\cdots \tau_2^{\nu_2}\rangle
\end{equation}
where the variables $\nu_j$ can assume values between $0$ and $k$ (subject
to the constraint $\nu_2+\nu_3+\cdots+\nu_n=k$).
In a population of constant size, $x=1$, the variables $\tau_{j}$ are independent and their correlation functions 
factorise. In general this is not the case:
\citet{Ziv:2008}, for example, 
have calculated $\langle \tau_i\tau_j\rangle$
for a smoothly varying population (Eqs.~(2) and (3) in their paper).

In the following we show how the 
correlation functions of arbitrary order appearing in (\ref{eq:Tnk2}) 
can be calculated in a very simple manner. 
Consider first the case $k=1$. We have
\begin{equation}
\label{eq:taujind}
\tau_j = \int_0^\infty{\rm d}t\, 1_{\{\ell(t)=j\}}\,.
\end{equation}
Here $\ell(t)$ denotes the number of lines for a particular realisation
of the coalescent process at time $t$ in a sample of size $n=\ell(0)$. The indicator
function in Eq.~(\ref{eq:taujind}) is unity when $\ell(t) = j$ and zero otherwise.
Averaging over realisations gives    
\begin{equation}
\langle \tau_j\rangle = \int_0^\infty{\rm d}t \langle 1_{\{\ell(t)=j\}}\rangle
= \int_0^\infty {\rm d}t f_{nj}(0,t)\,.
\end{equation}
Here $f_{nm}(t_1,t_2)$
is the conditional probability
that $n$ ancestral lines at $t_1$ coalesce to $m$
lines at time $t_2 > t_1$.

For a constant population size ($x=1$),
the coalescent is invariant
under time translations,
$f_{nm}(t_1,t_2) = g_{nm}(t_2-t_1) H(t_2-t_1)$. Here
$H(t)=1$ if $t > 0$ and zero otherwise.
The conditional probability $g_{nm}(t)$ was derived by \citet{Tav:84}.
For $m\geq 2$ the result is:
\begin{eqnarray}
\label{eq:gnm}
g_{nm}(t) &=& \sum_{j=m}^n c_{nmj} {\rm e}^{-b_j t}\\
c_{nmj} &=&
 (-1)^{j-m}
\frac{2j-1}{m!(j-m)!}
\frac{\Gamma(m+j-1)}{\Gamma(m)}
\frac{\Gamma(n)}{\Gamma(n+j)} \frac{\Gamma(n+1)}{\Gamma(n-j+1)} \,.
\end{eqnarray}
In the general case of a variable population size, as shown by \citet{GriTav:94}, the conditional probability
depends only on the intensity $\Lambda(t_2)-\Lambda(t_1)$ during
the time-interval $[t_1,t_2]$:
\begin{equation} 
f_{nm}(t_1,t_2) = g_{nm}\big(\Lambda(t_2)-\Lambda(t_1)\big)\,.
\end{equation}

Now consider the case $k=2$. For $i>j$ we have simply
\begin{eqnarray}
\label{eq:tij2}
\tau_i \tau _j &= & \int_0^\infty {\rm d}t_1  1_{\{\ell(t_1) = i\}}
 \int_{0}^\infty {\rm d}t_2 1_{\{\ell(t_2)=j\}}\\
&=& \int_0^\infty {\rm d}t_1 1_{\{\ell(t_1)=i\}} \int_{t_1}^\infty {\rm d}t_2 1_{\{\ell(t_2)=j\}}\,,
\nonumber
\end{eqnarray}
because the second indicator function vanishes when $t_2 < t_1$.
Averaging over realisations we find:
\begin{equation}
\label{eq:tij}
\langle \tau_i\tau_j\rangle = \int_0^\infty {\rm d}t_1 f_{ni}(0,t_1)
\int_{t_1}^\infty {\rm d}t_2 f_{ij}(t_1,t_2)\,.
\end{equation}
This result is illustrated in Fig. \ref{fig:theta}{\bf b}. In deriving
it we have used 
the multiplicative rule
\bal{ 
	\expt{1_{\{\ell(t_1)=i\}} 1_{\{\ell(t_2)=j\}}} = f_{ni}(0,t_1)f_{ij}(t_1,t_2).
}
For $i=j$, by contrast, we find 
\begin{eqnarray}
\tau_j^2 &=& \int_0^\infty {\rm d}t_1 1_{\{\ell(t_1)=j\}} \int_{0}^\infty {\rm d}t_2 1_{\{\ell(t_2)=j\}}\nonumber\\
&=& 2\int_0^\infty {\rm d}t_1  1_{\{\ell(t_1)=j\}}\int_{t_1}^\infty {\rm d}t_2 1_{\{\ell(t_2)=j\}}\,,
\end{eqnarray}
which upon averaging yields
\begin{equation}
\label{eq:tau2}
\langle \tau_j^2 \rangle = 2
 \int_0^\infty {\rm d}t_1 f_{nj}(0,t_1) \int_{t_1}^\infty {\rm d}t_2 f_{jj}(t_1,t_2)\,.
\end{equation}
More general correlation functions are readily obtained in terms of multiple integrals
over the functions $f_{nm}$.  Inserting into (\ref{eq:Tnk2}) we see that the
combinatorial factors $(\nu_2!)^{-1}$ $\cdots$ $ (\nu_n!)^{-1}$ cancel to obtain
\begin{eqnarray}
\label{eq:corr}
\langle T_n^k\rangle &=&k!  \sum_{m_1=2}^n \sum_{m_2=2}^{m_1} \!\cdots\! \sum_{m_k = 2}^{m_{k-1}} \!m_1 \cdots m_k 
\int_0^\infty\!\!\!\! \!{\rm d}t_1 
f_{nm_1}(0,t_1)
\cdots\!\!
\int_{t_{k-1}}^\infty\!\!\!\!\!{\rm d}{t_k} f_{m_{k-1}m_k}(t_{k-1},t_{k})\,.
\end{eqnarray}
Eq. (\ref{eq:corr}) provides an explicit expression for the moments of
the total branch lengths $T_n$ in populations with smooth population-size variations.

Note that Eq.~(\ref{eq:corr}) expresses the $k$-th moment of $T_n$ in terms of
a $2k$-fold sum (according to (\ref{eq:gnm}) each factor of $f_{n_im_i}$ contains a sum over $j_i$).
Eq.~(\ref{eq:corr})  can be further simplified by explicitly performing
the sums over $m_1,\ldots,m_k$. This results in
\begin{eqnarray}
\label{eq:corr2}
\langle T_n^k\rangle &=& k! \sum_{j_1=2}^n \cdots \sum_{j_k=2}^{j_{k-1}}
d_{n;j_1,\ldots,j_k}
\int_0^\infty\!\!\!\! \!{\rm d}t_1 {\rm e}^{-b_{j_1}\Lambda(t_1)}
\int_{t_{1}}^\infty\!\!\!\!\!{\rm d}{t_2} {\rm e}^{-b_{j_2}[\Lambda(t_2)-\Lambda(t_{1})]}\\\nonumber
&&\hspace*{2cm}\cdots\!\!
\int_{t_{k-1}}^\infty\!\!\!\!\!{\rm d}{t_k} {\rm e}^{-b_{j_k}[\Lambda(t_k)-\Lambda(t_{k-1})]}\,.
\end{eqnarray}
The coefficients are determined by recursion:
\begin{eqnarray}
\label{eq:d1}
d_{n;j} &=& \sum_{m=2}^{j} m \,c_{nmj} =  (2j-1) (1+(-1)^{j}) 
\frac{\binom{2n-1}{n-j}}{\binom{2n-1}{n}}
\,,\\
d_{n;j_1,\ldots, j_k} &=& \sum_{m=j_{2}}^{j_{1}} m\, c_{nmj_1} d_{m;j_2,\ldots j_k}\,.
\label{eq:d2}
\end{eqnarray}
For the first moment, an expression corresponding
to Eq.~(\ref{eq:corr}) for the particular case $k=1$ was derived
by \citet{Slatkin}. Evaluating (\ref{eq:corr}) for
$k=1$ using (\ref{eq:corr2}) and (\ref{eq:d1}) we find
\begin{equation}
\label{eq:Tn5}
\langle T_n\rangle = 
\frac{1}{\binom{2n-1}{n}} 
\sum_{i=1}^{\lfloor n/2\rfloor} (4i-1) \binom{2n-1}{n-2i} 
\int_0^\infty \!\!\!{\rm d}t\, {\rm e}^{-i(2i-1) \Lambda(t)}\,.
\end{equation}
Here $\lfloor\cdots\rfloor$ denotes taking the integer part.
This result is equivalent to Eq.~(3) in \citep{Aus:1997},
and also
to the result obtained by summing Eq.~(1) in \citep{Ziv:2008}.
For $k=2$, the coefficients $d_{n,;j_1,j_2}$ are tabulated in Tab. \ref{tab:1}
in appendix \ref{sec:B} for small values of $n$.
In general, the nested integrals in Eq.~(\ref{eq:corr2}) cannot be simplified further; their form
expresses the correlations of the times $\tau_j$ due to population-size variations.

Finally note that for $n=2$, Eq.~(\ref{eq:corr}) can be evaluated to give (\ref{eq:T2k}).
We show this explicitly because it demonstrates how the expression (\ref{eq:corr}) simplifies
when $k > n$. We have
\begin{eqnarray}\nonumber
\langle \tau_2^k\rangle &  =& k! \int_0^\infty {\rm d}t_1 f_{22}(0,t_1)
\int_{t_1}^\infty {\rm d}t_{2}  f_{22}(t_1, t_2)\cdots \int_{t_{k-1}}^\infty {\rm d}t_k f_{22}(t_{k-1},t_k)\\
&=&k! \int_0^\infty {\rm d}t_1 
\int_{t_1}^\infty {\rm d}t_{2}   \cdots \int_{t_{k-1}}^\infty {\rm d}t_k f_{22}(0,t_{k})
= k \int_0^\infty {\rm d} t \,t^{k-1} {\rm e}^{-\Lambda(t)}\,.
\end{eqnarray}
This yields Eq.~(\ref{eq:T2k}).

We conclude this section by remarking that
appendix \ref{sec:A} summarises an alternative approach to calculating $F_n(q)$ and the moments
 $\langle T_n^k\rangle$, again resulting in Eqs.~(\ref{eq:r2}) and (\ref{eq:corr}). The approach
 described in appendix \ref{sec:A} yields  a simple
 recursion, Eq.~(\ref{eq:mom_9}),  which allows for a convenient calculation
 of the moments $\langle T_n^k\rangle$.
This result also demonstrates explicitly how the moments $\langle T_n^k\rangle$,
for a given  curve $x(t)$, depend upon the time at which the population is sampled.

In the following two sections we describe a simple population model
subject to population-size variations, and compare results of numerical
simulations of this model to the analytical results obtained above.

\section{A model for a population with time-dependent carrying capacity}
\label{sec:sim}
The purpose of this section is to describe a modified Wright-Fisher model with fluctuating
population size. This model is used in the numerical simulations of sample genealogies
described in Sec. \ref{sec:num}. Recall the three key assumptions
of the Wright-Fisher model: (a) constant population-size, (b) discrete, non-overlapping generations,
(c) a symmetric multinomial distribution of family sizes.
We have adopted the following approach: in our simulations, assumptions (b) and
(c) are still satisfied, but assumption (a) is relaxed.

We study  a large but finite population of fluctuating size $N_\tau$, where $\tau=1,2,\ldots$
labels the discrete, non-overlapping generations forward in time. 
The model we have adopted is
the following:
consider a generation $\tau$ consisting
of $N_\tau$ individuals. The number of individuals in generation $\tau+1$
is then given by
\begin{equation}
N_{\tau+1} = \sum_{j=1}^{N_\tau} \xi_j
\end{equation}
where the random family sizes $\xi_j$ are 
independent and identically distributed
random variables having a Poisson distribution with parameter $\lambda_\tau$ (specified
below). Consequently the number $N_{\tau+1}$ is Poisson distributed
with mean $N_\tau \lambda_\tau$. 

This model exhibits a fluctuating population size $N_\tau$, rapidly changing from generation to generation. As pointed out in the introduction, in large populations such fluctuations are averaged over
by the ancestral coalescent process, and can be captured in terms of an effective population size.
The resulting genealogies are simply described by Kingman's coalescent for a constant effective
population size of the form (\ref{eq:Neff}).

Interesting population-size fluctuations occur on larger time scales, corresponding to \lq slow' variations
of the population size over several generations. Such slow changes are most commonly  interpreted as 
consequences of a changing environment. A natural model for such changes is to impose
a finite carrying capacity $K_\tau$ on the population which varies as a function of $\tau$.
This is the approach adopted in the following, and we choose
\bal{
\label{eq:lt}
	\lambda_\tau = \frac{1+r}{1 + r\, N_\tau/K_{\tau+1}}
}
for a certain parameter value $r > 0$. Here $K_{\tau+1}$ is the carrying capacity in generation $\tau+1$. 
If the environmental changes affected the population
through fertility variations,  $K_{\tau+1}$ would be replaced by  $K_{\tau}$ in Eq.~(\ref{eq:lt}).
Eq.~(\ref{eq:lt}) is chosen so that the population ceases to
grow on average when the carrying capacity is reached ($\lambda_\tau=1$ for $N_\tau=K_{\tau+1}$). 
When the population size is small, the population growth follows the logistic law,
$\lambda_\tau = 1+r(1-N_\tau/K_{\tau+1})$, where $r$ is the logistic growth rate. 
The particular form of  Eq.~(\ref{eq:lt}) ensures that $\lambda_\tau > 0$ .

Note that fluctuations of $N_\tau$ in this model are due to two different sources: rapid fluctuations 
are caused by the randomness of the family sizes, slow fluctuations are caused by the time dependence
of the carrying capacity. Our choice for the time dependence of $K_\tau$ 
is dictated by the following considerations. 
The aim is to describe the influence
of a fluctuating population size upon the statistics of genetic variation. 
To this end we need to consider the functional form of $K_\tau$.  A simple 
choice for $K_\tau$ is a periodically varying function, such as
\begin{equation}
\label{eq:Ktau}
K_\tau = K_0[1 + \epsilon\sin( 2 \pi \nu \tau)]\,.
\end{equation}
Note that a more complex dependence of $K_\tau$ upon $\tau$ can be obtained from superpositions of such functions
with different amplitudes $\epsilon$ and frequencies $\nu$.
Here we use simply (\ref{eq:Ktau}), and investigate how the statistics
of genetic variation in a sample depends upon frequency of the fluctuations
in $K_\tau$.

\begin{figure}[t]
\centerline{\includegraphics[width=8cm]{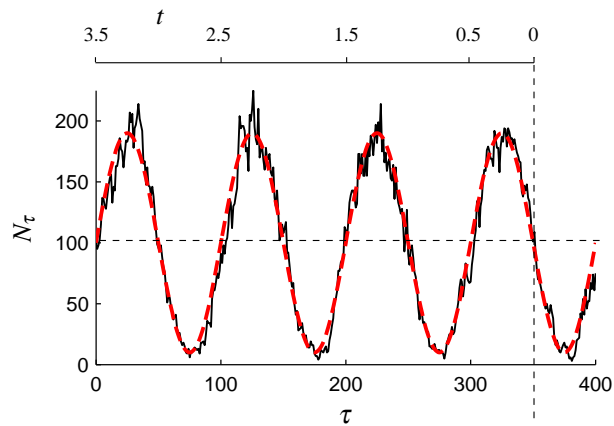}}
\caption{\label{fig:Nt} 
Shows one realisation of the curve $N_\tau$ obtained from simulations of the model described in Sec. \ref{sec:sim} (black solid line). 
Choice of parameters: $r = 1$, $K_0 = 100$, $\epsilon = 0.9$, and $K_0\nu = 1$.
Also shown is an average over the fast fluctuations, 
$K_0 [1+\epsilon\sin(2\pi \nu \tau)]$, red dashed line.
The upper horizontal axis illustrates where the population is sampled, and how time is counted backwards in
the coalescent approximation. $N_0$ denotes the size of the population at the time of sampling.}
\end{figure}

Fig. \ref{fig:Nt} shows a realisation of a curve $N_\tau$ obtained in this manner (the choice of parameters is given in the figure caption).
The figure clearly exhibits fluctuations in $N_\tau$ on two time scales. As pointed out above, we are interested
in determining the effect of the size variations occurring at long time scales.

Last but not least 
we note that conditional on the sequence of population sizes, the genealogy of a set of individuals sampled at time $\tau$ 
can be determined recursively by randomly choosing ancestors in the preceding generations. This is ensured by the assumption 
that, 
 conditioned on the values of $N_\tau$ and $N_{\tau+1}$, 
the family sizes follow a symmetric multinomial distribution
$\mbox{Mn}(N_{\tau+1};\frac{1}{N_\tau},\ldots,\frac{1}{N_\tau})$.
The resulting correspondence with the Wright-Fisher rule of reproduction ensures
that the genealogies can be determined recursively in the way suggested above.

\section{Comparison between numerical simulations and coalescent predictions}
\label{sec:num}
In this section we discuss the numerically
computed distributions shown in Fig.~\ref{fig:dist}
in terms of the results obtained using  the coalescent
approximation. 
The shapes observed in Fig.~\ref{fig:dist}
are conveniently characterised in terms their mean $\langle T_n\rangle$, 
variance, skewness, and kurtosis:
\begin{align}
\mbox{var}(T_n) &= \langle T_n^2\rangle - \expt{T_n}^2\,,\nonumber\\
\label{eq:skew}
\mbox{skew}(T_n) &= \frac{\expt{(T_n - \expt{T_n})^3}}{\mbox{var}^{3/2}(T_n)}\,,\\
\mbox{kurt}(T_n) &= \frac{\expt{(T_n - \expt{T_n})^4}}{\mbox{var}^2(T_n)}\,.
\end{align}
Recall that for a normal distribution the skewness
vanishes, and the kurtosis equals three.
We can write the skewness and kurtosis in terms of the moments $\expt{T_n^k}$ using
$\expt{(T_n - \expt{T_n})^3} = \langle T_n^3\rangle - 3 \langle T_n^2\rangle\langle T_n\rangle + 2\langle T_n\rangle^3$ 
and  
$\expt{(T_n - \expt{T_n})^4} =\langle T_n^4\rangle-4\langle T_n^3\rangle\langle T_n\rangle + 6 \langle T_n^2\rangle\langle T_n\rangle^2 - 3 \langle T_n\rangle^4$ .

As argued in Sec. \ref{sec:sim} and as shown in Fig. \ref{fig:Nt}, our model populations exhibit fast size changes due to the random
distribution of family sizes. 
As pointed out in the introduction,
these fluctuations are averaged over by the genealogical process and need not be considered.
The model populations are also subject to slow (and deterministic) size fluctuations given by the
time-dependence (\ref{eq:Ktau}) of the carrying capacity. Averaging over the fast fluctuations 
these give rise to a smooth population-size dependence $x(t)$. Given Eq. (\ref{eq:Ktau}), 
the distribution of $T_n$ depends upon the instance in time when the population is sampled.
In the simulations we sampled at a particular point (illustrated in Fig. \ref{fig:Nt} as a dashed
vertical line), so that
\begin{equation}
\label{eq:xt}
x(t) = 1+\epsilon \sin(\omega t)\,.
\end{equation}
Here the frequency is given by $\omega = 2\pi \nu K_0$, and time $t$
is now counted backwards, as in Sec. \ref{sec:rec}. If the population
were sampled at a different time, the distribution $\rho(T_n)$ of $T_n$ (and hence
its moments and the corresponding function $F_n(q)$) would change: the distribution
depends  for example upon whether most recently the population was expanding or declining.
 The results derived in appendix \ref{sec:A} make it possible to determine the corresponding changes 
 to $\rho(T_n)$ in a transparent manner, but we do not discuss this issue further here.

\begin{figure*}[t]
\centerline{\includegraphics[width=16cm]{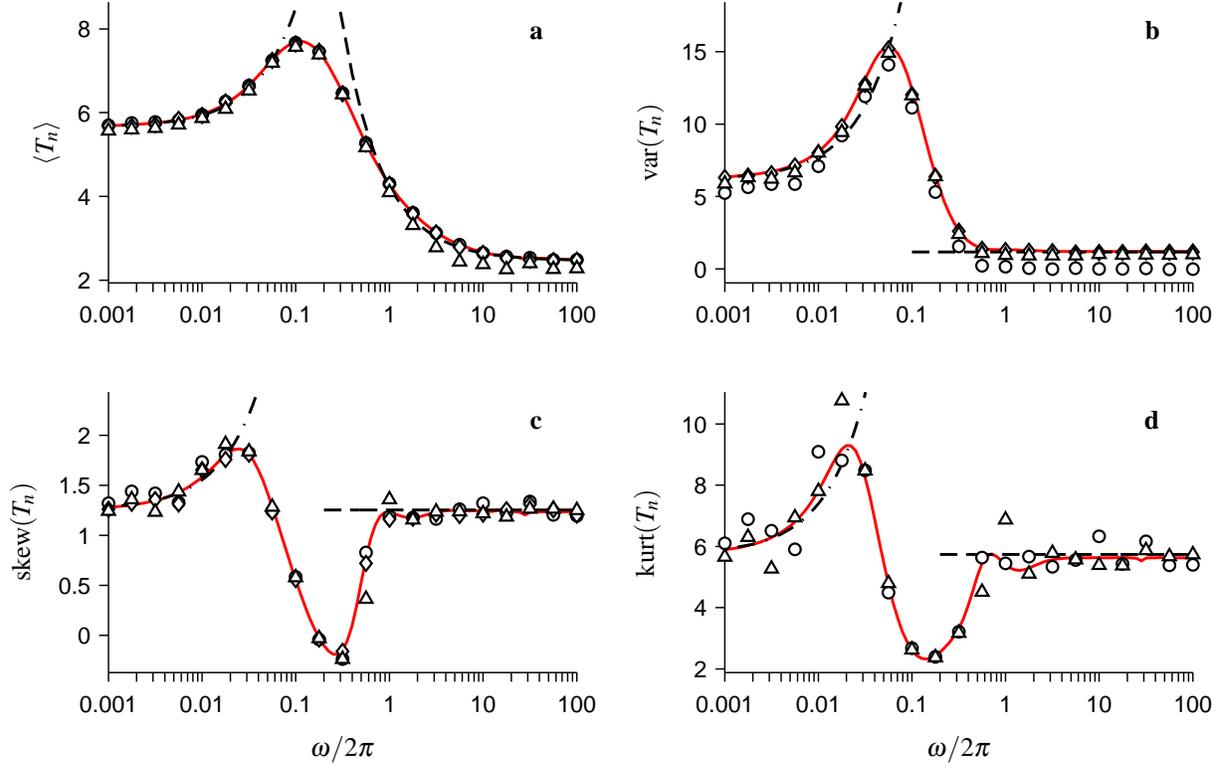}}
\caption{\label{fig:stoch_moments} 
Shows mean ({\bf a}), variance ({\bf b}), skewness ({\bf c}) and kurtosis ({\bf d}) of the distribution of $T_n$ for samples of size $n=10$, as a function of the frequency of the population-size fluctuations. Shown are results of numerical simulations ($10,000$ simulations, $K_0 = 100$, triangles; $K_0 = 1,000$, diamonds; and $K_0 = 10,000$, circles) as well as results 
computed within the coalescent approximation described
in Sec. \ref{sec:rec}, red solid lines. Black dash-dotted 
and dashed lines show
 the approximations for small frequencies, 
Eqs. (\ref{eq:Tn6}) and (\ref{eq:vn}),
and for large frequencies, Eqs. (\ref{eq:Tn7}) and (\ref{eq:Tn11}).
The expressions for the limiting behaviours of the skewness and the kurtosis
are shown in panels {\bf c} and {\bf d}, but are not given in the text.
The remaining parameter values are $r = 1$ and $\epsilon = 0.9$, as in Fig. \ref{fig:dist}.
}
\end{figure*}

Fig.~\ref{fig:stoch_moments} shows how the mean, 
variance, skewness, and kurtosis of
the distribution of $T_n$ depend on the  frequency $\omega$  of
the population size variation, Eq. (\ref{eq:xt}). Shown are results of numerical simulations
of the model described in section \ref{sec:sim} (symbols), and results
obtained within the coalescent approximation using Eq.~(\ref{eq:mom_9}).
We observe that the coalescent approximation describes the results of the
numerical simulations well, even for 
small population sizes.
 
In the numerical simulations we have found that, for very small population 
sizes, random fluctuations of $N_\tau$ around the time-dependent 
carrying capacity $K_\tau$ become increasingly important. Since we
suspected that the small deviations observed in Fig. \ref{fig:stoch_moments}{\bf a}
for $K_0=100$ were due to such fluctuations, we performed slightly
modified simulations imposing a deterministic law upon $N_\tau$ by forcing
$N_\tau = K_\tau$ in every generation (where $K_\tau$ is given by
(\ref{eq:Ktau})). 
Comparison of the corresponding results (not shown) with 
Fig.~\ref{fig:stoch_moments}{\bf a} indicates that the deviations for $K_0=100$ at large frequencies are indeed caused
 by the stochastic fluctuations in the population dynamics underlying Fig.~\ref{fig:stoch_moments}{\bf a}.
A different interpretation of this effect is the following: when the population size is very small, and when $\epsilon$ is close to unity,
the population may exhibit a non-negligible probability of becoming 
extinct during the expected time to the most recent common ancestor
for a sample of size $n$.
In this case we have conditioned on the existence of the population 
during $100 K_0$ generations using rejection sampling. In practice
this avoids extinction, but it leads to a 
biased size distribution.

Consider now  the frequency dependence of the moments shown 
in Fig. \ref{fig:stoch_moments}. It can be qualitatively and quantitatively
understood using Eq. (\ref{eq:corr2}) together with the following
expression for $\Lambda(t)$:
\begin{align}
\label{eq:Lt}
\Lambda(t) = \int_0^t\!\!\!\frac{{\rm d}s}{1+\epsilon \sin(\omega s)}=
\frac{
 \Big\lfloor \frac{\omega t}{2\pi}+\frac{1}{2}\Big\rfloor-\frac{1}{\pi}
\arctan\Big(\frac{\epsilon}{\sqrt{1-\epsilon^2}}\Big)
+\frac{1}{\pi}\arctan\Big(\frac{\tan(\omega t/2)+\epsilon}%
{\sqrt{1-\epsilon^2}}
\Big) }{(\omega/2\pi)\sqrt{1-\epsilon^2}}\,.
\end{align}
We discuss the limits of small and large frequencies $\omega$
separately.
In the limit of $\omega\rightarrow 0$, Eq.~(\ref{eq:Lt}) 
simplifies to $\Lambda(t) \approx t -\frac{1}{2} \epsilon\omega t^2$ .
Inserting this into (\ref{eq:Tn5}) and approximating 
\begin{equation}
\int_0^\infty\!\!\!\!{\rm d} t\, {\rm e}^{-b_j \Lambda(t)}
\approx \frac{1}{b_j}\left(1+\frac{\epsilon\omega}{b_j}\right)
\end{equation}
we find 
\begin{equation}
\label{eq:Tn6}
\langle T_n \rangle \approx 2h_n 
+4\epsilon\omega(g_n-h_n/n)\,.
\end{equation}
Here $h_n = \sum_{j=1}^{n-1} j^{-1}$ and   $g_n = \sum_{j=1}^{n-1} {j^{-2}}$.
Eq.~(\ref{eq:Tn6}) is
shown in Fig. \ref{fig:stoch_moments}{\bf a} as a dash-dotted line.
To compute the variance we approximate
\begin{equation}
\int_0^\infty\!\!\!{\rm d}t_1 {\rm e}^{-b_{j_1}\Lambda(t_1)}\int_{t_{1}}^\infty\!\!\!\!\!{\rm d}{t_2} {\rm e}^{-b_{j_2}[\Lambda(t_2)-\Lambda(t_{1})]}
\approx\frac{1}{b_{j_1}b_{j_2}}+\epsilon\omega \frac{b_{j_1}+2b_{j_2}}{b_{j_1}^2b_{j_2}^2}\,,
\end{equation}
and find an approximate expression for $\langle T_n^2 \rangle$ which
results in the following expression for the variance:
\begin{equation}
\label{eq:vn}
 \mbox{var}(T_n) 
\approx 4 g_n + 16 \epsilon\omega \big(f_n-g_n+\frac{h_n-g_n}{n}\big)
\end{equation} 
with $f_n = \sum_{m=1}^{n-1}({m^{-3}}+ m^{-2}{h_{m+1}})$.
The limiting value for zero frequency is that of the standard coalescent with constant
population size $x=1$.
Eq. (\ref{eq:vn}) is shown in Fig. \ref{fig:stoch_moments}{\bf b} as a dash-dotted line.
Similarly the standard results for the constant-size
coalescent are obtained for the skewness and for the kurtosis
in the limit of $\omega \rightarrow 0$.
This limiting behaviour is illustrated in Fig. \ref{fig:dist}{\bf a}
which shows that the distribution of $T_n$ approaches
that for Kingman's coalescent for a constant population size
$x=1$ in the limit of small frequencies. We note that for $\omega \ll 1$,
the population-size dependence is essentially that of a declining population, because
the time to the most recent common ancestor is reached before the first maximum in $x(t)$
going backwards in time (see Fig. \ref{fig:Nt} and Eq. (\ref{eq:xt}). 

Of particular interest is the limit of large frequencies, as we now show.
As the frequency tends to infinity, 
one expects that the coalescent
process averages over the population-size oscillations, and the standard
coalescent process with a constant effective population size should be obtained.
For large but finite frequencies, by contrast, Fig. \ref{fig:stoch_moments}{\bf a}
exhibits deviations from the standard coalescent behaviour. 
In the following we analyse the  behaviour of the moments in this regime.
In the limit of large frequencies,
Eq.~(\ref{eq:Lt}) simplifies to
\begin{eqnarray}
\label{eq:Lt2}
\Lambda(t) &=& \frac{t}{\sqrt{1-\epsilon^2}}
-\frac{\arctan\Big(\frac{\epsilon}{\sqrt{1-\epsilon^2}}\Big)}
{\omega\sqrt{1-\epsilon^2}}+ O(\omega^{-2})+\mbox {oscillatory terms}\,.
\end{eqnarray}
For large frequencies, the function $\Lambda(t)$ is well approximated
by a shifted linear function
\begin{equation}
\label{eq:L0}
\Lambda(t) \approx t/x_{\rm eff}  + \overline{\Lambda}_0\,.
\end{equation}
Here 
\begin{equation}
x_{\rm eff} =
 \left(\lim_{T\rightarrow\infty}\frac{1}{T} \int_0^T\frac{{\rm d} t}
{1+\epsilon\sin(\omega t)}\right)^{-1}
= \sqrt{1-\epsilon^2}
\end{equation}
is the effective population size according to Eq.~(\ref{eq:Neff}), it describes
the influence of the demographic fluctuations upon the part of the genealogy
in the far past. The small offset
\begin{equation}
\overline{\Lambda}_0 = -\frac{\arctan(\epsilon/\sqrt{1-\epsilon^2})}{x_{\rm eff} \omega}\approx -\frac{\epsilon}{x_{\rm eff} \omega}
\quad\mbox{for $\epsilon$ not too close to unity}
\end{equation}
describes the influence of demographic changes on the most recent part of the genealogy.
Inserting the approximation (\ref{eq:L0}) into (\ref{eq:corr2}) we find for large
frequencies (and when the  amplitude $\epsilon$ is not too close to unity):
\begin{equation}
\label{eq:Tn7}
\langle T_n\rangle
\approx  2x_{\rm eff}h_n + \frac{n\,\epsilon}{\omega}\,.
\end{equation}
The first term in (\ref{eq:Tn7}) is the expected time of Kingman's coalescent
for a constant effective population size $x_{\rm eff}$.
The curve corresponding to (\ref{eq:Tn7}) is shown as a dashed line
in Fig. \ref{fig:stoch_moments}{\bf a}.  

We now discuss the behaviour of  the variance
shown in Fig. \ref{fig:stoch_moments}{\bf b}. For the second moment we find:
\begin{eqnarray}
\label{eq:Tn11}
\langle T_n^2\rangle
&\approx&  4 x_{\rm eff}^2 (g_n+h_n^2)+
\frac{4nh_nx_{\rm eff}\epsilon}{\omega} \,,
\end{eqnarray}
The first term in Eq. (\ref{eq:Tn11}) corresponds to the second
moment of $T_n$ in Kingman's coalescent
with a constant effective population size $x_{\rm eff}$.
The second term in (\ref{eq:Tn11}) represents
a correction due to finite but large
frequencies, it depends in a simple fashion
on the effective population size $x_{\rm eff}$
and on the sample size $n$. 

Comparing
Eqs. (\ref{eq:Tn7}) and (\ref{eq:Tn11})
we arrive at the conclusion that the corresponding correction
for the variance $\mbox{var}(T_n)$ vanishes.
This is consistent with the fact that, 
at large frequencies, the variance of $T_n$
is surprisingly insensitive to changes in frequency (as opposed to the 
behaviour of $\langle T_n\rangle$, see Fig. \ref{fig:stoch_moments}{\bf a} and {\bf b}).
In fact, the limiting value (shown in Fig. \ref{fig:stoch_moments}{\bf b}
as a dashed line) is a very good approximation to $\mbox{var}(T_n)$
down to $\omega \approx 3$.

Consider now the
 skewness and the kurtosis shown in Figs. \ref{fig:stoch_moments}{\bf c}
and {\bf d}. Their behaviour is similar to that of the variance: over a substantial
range, the skewness and the kurtosis  are essentially independent of $\omega$.
The results shown in Fig. \ref{fig:stoch_moments} imply
that over a large range of frequencies, 
the distribution of the total branch lengths $T_n$  can be approximated
as follows: 
the distribution is essentially that of the standard
Kingman coalescent with an effective population size $x_{\rm eff}$, but the distribution
is shifted such that its mean is given by Eq. (\ref{eq:Tn7}), rather than by
$2 x_{\rm eff} h_n$.

One may wonder when this \lq rigid shift' occurs. Given Eq. (\ref{eq:corr})
it is straightforward to work out the fluctuations of the times $\tau_j$ within the approximation
(\ref{eq:L0}). We find that for $j < n$, the expected value of $\tau_j$ is exactly that
of the standard Kingman coalescent with effective population size $x_{\rm eff}$.
But for $j=n$ it is rigidly shifted by $-x_{\rm eff} \overline{\Lambda}_0$.
This indicates that the genealogies are essentially those of the standard coalescent, but
modified by an initial rigid shift. In the parameter regime discussed here, the distribution
of times is expected to be well approximated by a two-parameter family of distributions:
\bal{\label{eq:large_w_approx}
       P(T_n < z) \approx \bigg[1 - \exp\!\Big(\!-\frac{z/x_{\rm eff} + n\overline{\Lambda}_0}{2}\Big)\bigg]^{n-1}
}
when $z/x_{\rm eff} > -n \overline{\Lambda}_0$, and $P(T_n < z) \approx 0$ for smaller values of $z$.
The first parameter is the effective population size $x_{\rm eff}$ which determines
the slope of the function $\Lambda(t)$ at large times and describes the demographic
effect on the far past of the genealogy. The second parameter, $\overline{\Lambda}_0$
describes the influence of the demographic fluctuations on the initial
part of the sample genealogy.  This parameter can be negative (initial population expansion, this is the case shown in Fig. \ref{fig:stoch_moments}) 
or positive (initial population decline). When $\overline{\Lambda}_0 > 0$,
the distribution $\rho(T_n)$ is rigidly shifted to the left. In this case
the approximation (\ref{eq:L0}) is expected to break down when the body
of the distribution reaches $T_n=0$.

Note that the distribution (\ref{eq:large_w_approx})
cannot be described by a single parameter (a \lq generalised effective population size').
The approximation (\ref{eq:large_w_approx}) was used to generate
the red dashed curves in Fig. \ref{fig:dist}{\bf d} and {\bf e}.

\section{Discussion and conclusions}
\label{sec:conclusions}
The aim of this paper was to investigate how the frequency of smooth
population-size fluctuations determines the shape of the distribution
of total branch lengths of sample genealogies, and thus of statistical
measures of genetic variation.

We have performed simulations for a modified Wright-Fisher
model of a population subject to a time-periodically varying carrying capacity
and have determined the distribution of the total branch lengths, shown
in Fig.~\ref{fig:dist}. We have characterised how the shapes
of the distributions depend upon the frequency of the population
size fluctuations by computing the frequency dependence of the
moments of these distributions. We could explain these
dependencies in terms of coalescent approximations.
In particular, we derived a general expression -- Eq. (\ref{eq:corr2}) --
for the moments $\langle T_n^k\rangle$ in populations
subject to smooth population changes of otherwise arbitrary
form.

Our results show how quickly (or slowly) the standard coalescent result
for a constant (effective) population sizes is recovered in
the limits of large and small frequencies. More importantly, 
our coalescent results allow to determine how significant
deviations are at large but finite frequencies. In this case we have 
argued that
at large frequencies,
the distribution of $T_n$ is essentially that of the standard Kingman coalescent
with an effective population size $x_{\rm eff}$, but with a shifted mean value
\begin{equation}
\langle T_n\rangle =  2x_{\rm eff} \sum_{j=1}^{n-1}\frac{1}{j} \,\,+ \frac{n\epsilon}{\omega}\,.
\end{equation}
The first term on the rhs corresponds to the result 
of the standard Kingman coalescent
with constant  effective population size $x_{\rm eff}$. The second
term on the rhs is the correction term resulting from the population-size
variations ($\epsilon$ is the amplitude
of the population-size oscillations, $\omega$ its frequency, and $n$ is the sample size).
Last but not least we have found that the coalescent approximation yields a reliable description of
the numerical data, even for very small populations.  

\begin{figure*}[t]
 \centerline{\includegraphics[width=16cm]{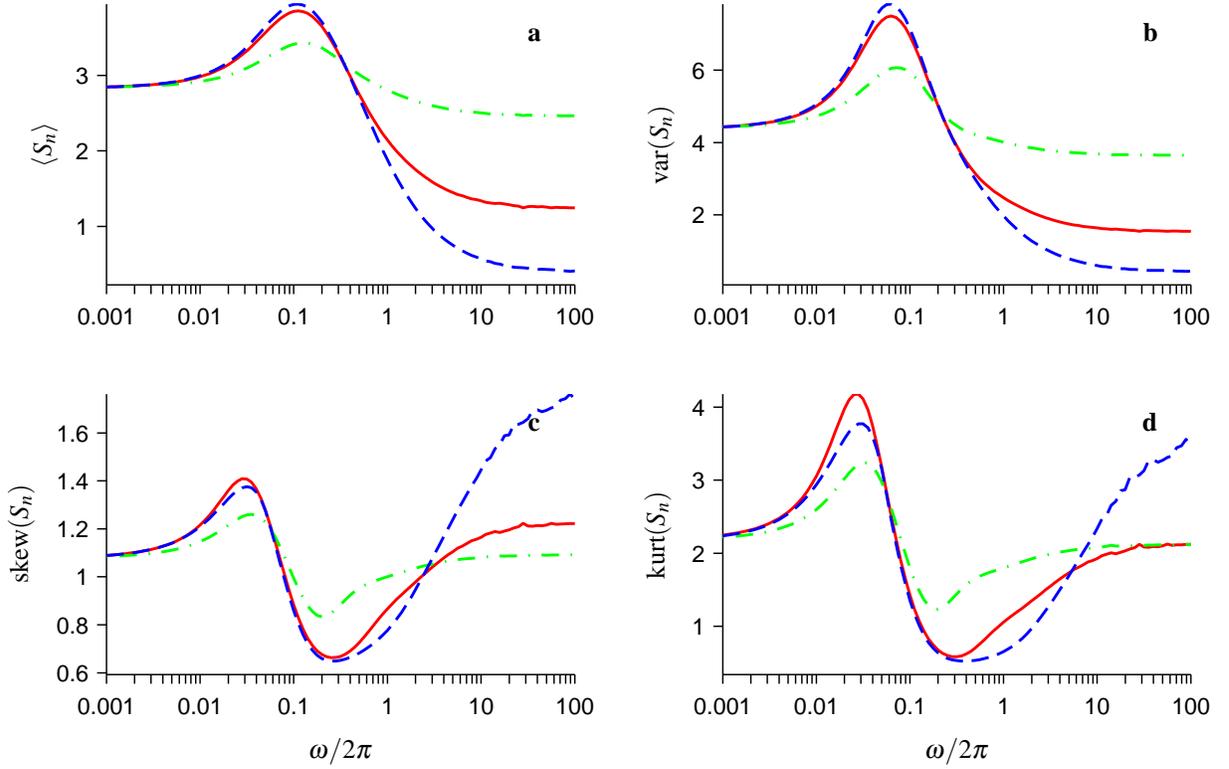}}
 \caption{\label{fig:sn1}
Shows mean ({\bf a}), variance ({\bf b}), 
skewness ({\bf c}) and kurtosis ({\bf d}) of the distribution of $S_n$ 
for samples of size $n=10$ and scaled mutation parameter $\theta=1$,
 as a function of the frequency of the population-size fluctuations, 
for three values of $\epsilon$: $\epsilon = 0.5$ (dashed-dotted green lines), 
$\epsilon = 0.9$ (solid red lines), and $\epsilon = 0.99$ (dashed blue lines). The curves were obtained by iteration of Eq. (\ref{eq:mom_9}) in combination with (\ref{eq:Lt}). 
}
\end{figure*}

\begin{figure*}[t]
 \centerline{\includegraphics[width=16cm]{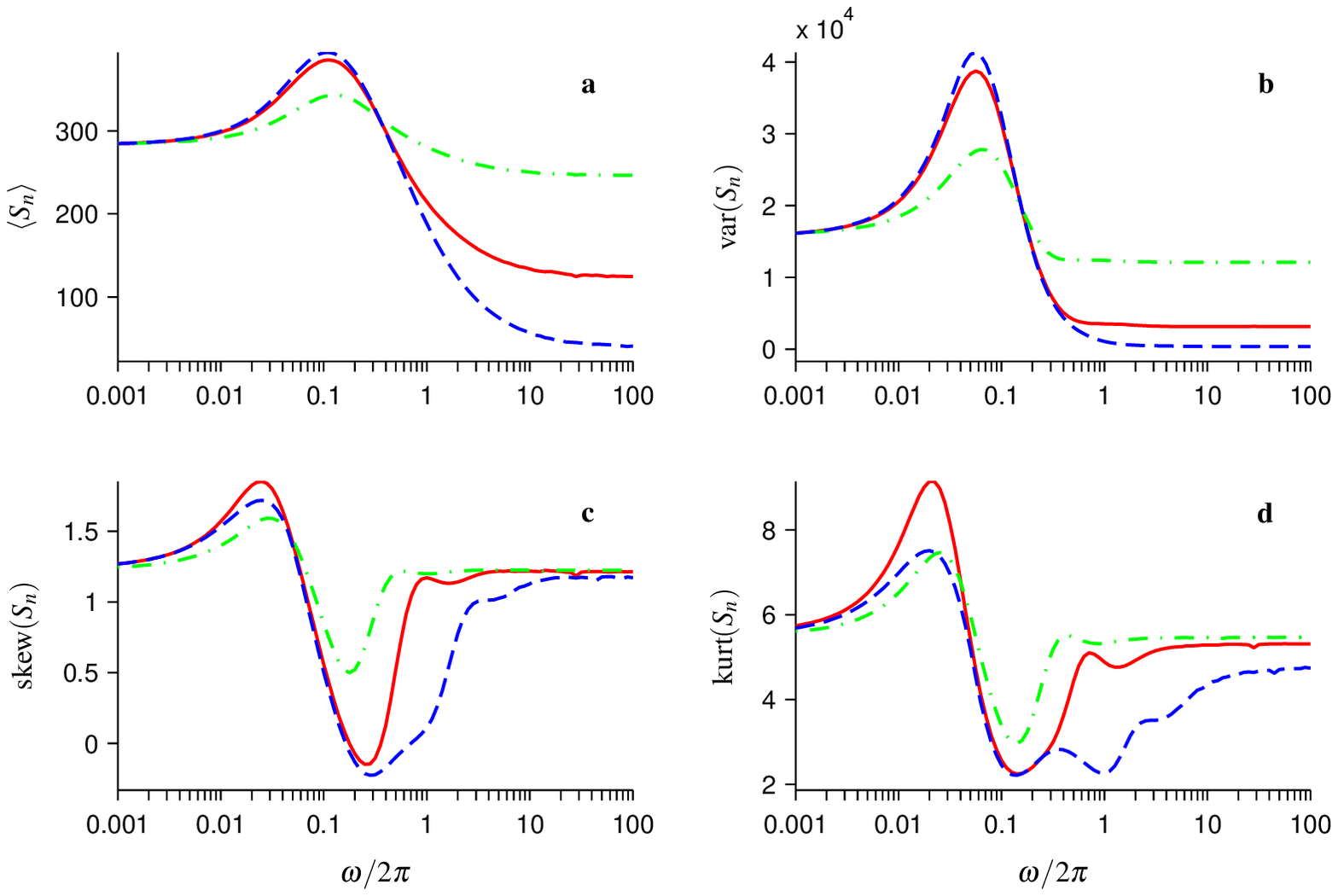}}
 \caption{\label{fig:sn100}
Same as Fig. \ref{fig:sn1}, but for $\theta = 100$. }
\end{figure*}

These results enable us to determine
how the distribution of the number $S_n$ of mutations
(single-nucleotide polymorphisms)
in a sample of size $n$ depends upon the frequency
and on the amplitude of population-size fluctuations:
Eq. (\ref{eq:pSn}) allows to compute moments of $S_n$
from Eq. (\ref{eq:corr2}). In this way we have determined
the mean, variance, skewness, and the kurtosis
of the distribution of $S_n$. The results
are shown in Figs. \ref{fig:sn1} and \ref{fig:sn100}.
As expected, the moments of $S_n$ approach those of $\theta T_n/2$ as the scaled mutation
parameter $\theta$ increases. This can be verified by comparing the red curves (corresponding
to $\epsilon = 0.9$) in Fig.~\ref{fig:sn100} to the red curves in Fig.~\ref{fig:stoch_moments}.
The higher moments converge more slowly than the mean and the variance. In conclusion, 
Figs. \ref{fig:sn1} and \ref{fig:sn100} demonstrate that the distribution 
of the number $S_n$ of mutations in samples of size $n$ depends in a complex manner
on the amplitude and on the frequency of the population-size variations, and
on the mutation parameter $\theta$. 

We close with a number of remarks. First, Eq.~(\ref{eq:corr2}) is easily generalised to describe
 the moments of observables which are polynomial functions of the times $\tau_j$ (see Fig. \ref{fig:theta}{\bf a} for
a definition of these times). Particularly simple
 is the case of observables $A$ that are linear functions of the times $\tau_j$,
 $A_n = \sum_{j=2}^n a_j \tau_j$. In this case the $k$-th moment
 of $A_n$ is given by Eq.~(\ref{eq:corr2}), but with
 modified coefficients: the factors $m$ in Eqs.~(\ref{eq:d1}) and (\ref{eq:d2})
 are replaced by $a_m$.

Second, some observables (such as the F-statistic \citep{FuLi:1993})
can be written as linear functions of $\tau_j$, but with random
coefficients. In this case too it is possible to explicitly
compute the moments of the distribution of the observable.
These two questions are addressed in a separate
paper \citep{Serik:2010}.

Third, a result derived in appendix \ref{sec:A}, 
Eq.~(\ref{eq:mom_9}), allows us to determine in a
transparent fashion how the fluctuations of $T_n$ and other
observables depend upon the time at which the population is sampled.
This will make it possible to discuss for example how Tajima's $D$-statistic
or  the $F$-statistic depend upon the time of sampling after
a bottleneck, a population expansion, or a decline.

Fourth, population-size fluctuations are sampled non-uniformly by the 
genealogies: initial coalescent events occur at faster rates and are thus  more sensitive to 
 recent size fluctuations. Remote coalescent events, by contrast, occur at  slower rates thus damping the effect of size fluctuations in the far past.
We therefore expect significant deviations from the standard coalescent behaviour arising
from the most recent history for large sample sizes $n$.
It would be interesting to quantify this expectation by computing
the covariances and higher moments of the times $\tau_j$ during which the sample
genealogy has $j$ lines: first for large $i\approx n$ and $j\approx n$ we expect to 
observe strong correlations $\langle \tau_i\tau_j\rangle - \langle 
\tau_i\rangle \langle \tau_j\rangle$ and thus deviations from the coalescent. Second for small values of  $i$ and $j$ we expect the times $\tau_i$ and $\tau_j$ to de-correlate and to follow
the distribution of the standard coalescent (with an effective 
population size).

Fifth, the model introduced in Sec. \ref{sec:sim} assumes a carrying capacity that varies sinusoidally, with a single frequency. 
It turns out, however, that our findings are valid for
arbitrary time-dependent fluctuations with sufficiently strong modes at small frequencies. Examples are linear combinations of high-frequency oscillations, or 
 stochastic fluctuations around a constant population size with sufficiently short auto-correlation time. 
In this more general case, too, we expect that $\Lambda(t)$ is
well approximated by (\ref{eq:L0}). If this is the case, the distribution
of times is of the form (\ref{eq:large_w_approx}) when $\overline{\Lambda}_0$ is small.

Taken together, the results derived in this paper give a rather
complete understanding of the fluctuations of empirical observables
due to smooth population size variations. These results will be
significant when attempting to disentangle the effects of
population-size variations from other factors influencing
genetic variation.

Our results raise the question under which circumstances the deviations from standard coalescent behaviours due to population-size fluctuations (Figs. \ref{fig:dist}, \ref{fig:stoch_moments}, \ref{fig:sn1}, and \ref{fig:sn100}) are most likely to strongly affect the interpretation of empirical data.  As our analysis indicates, 
the deviations become substantial when the frequency $\omega = 2\pi \nu K_0$ is of the order
of or less than the inverse expected time between coalescent events in the sample. Here $\nu$ is the frequency of the population size variations,
Eq. (\ref{eq:xt}), and $K_0$ is a suitable
measure of the population size (the arithmetically averaged carrying capacity in our example). 
In other words, rapid population-size fluctuations will have the strongest effect (other than
simply determining the effective population size, Eq. (\ref{eq:Neff})) in 
small local sub-populations with restricted gene flow between sub-populations with different fluctuations.  
The deviations are expect to be smaller at larger spatial scales, because the ancestral
process averages over the spatial fluctuations.
More generally, we conclude that deviations from standard coalescent behaviour are expected
for populations subject to an environment which smoothly changes as a function of space and time.
An example for such a population is the marine snail {\em L. saxatilis}. Its habitat
on the Northern coast of Bohusl\"an (Sweden) is fragmented into sub-populations with
strongly restricted gene flow between them, effective population sizes of sub populations
have been found to be very small \citep{Johannesson:2009}. Starting from the results derived
in this paper, we hope to determine gene genealogies
in such fragmented populations subject to smooth variations of population size in space and time.

{\em Acknowledgements}. Support from Vetenskapsradet,
The Bank of Sweden Tercentenary Foundation,
and from the Centre for Theoretical Biology
at the University of Gothenburg are gratefully acknowledged.

\bibliographystyle{genetics}

\begin{thebibliography}{25}
\expandafter\ifx\csname natexlab\endcsname\relax\def\natexlab#1{#1}\fi

\bibitem[{{\sc Austerlitz} {\em et~al.\/}(1997){\sc Austerlitz}, {\sc
  Jung-Muller}, {\sc Godelle} and {\sc Gouyon}}]{Aus:1997}
{\sc Austerlitz, B.}, {\sc B.~Jung-Muller}, {\sc B.~Godelle}, and {\sc
  P.~Gouyon}, 1997 Evolution of coalescence times, genetic diversity and
  structure during colonization.
\newblock Theor. Pop. Biol. {\bf 51}: 148--164.

\bibitem[{{\sc Ewens}(1982)}]{ewe82:neff}
{\sc Ewens, W.}, 1982 The concept of the effective population size.
\newblock Theor. Popul. Biol. {\bf 21}: 373--378.

\bibitem[{{\sc Fisher}(1930)}]{Fisher:1930}
{\sc Fisher, R.~A.}, 1930 {\em The genetical theory of natural selection.\/}.
\newblock Clarendon, Oxford.

\bibitem[{{\sc Fu} and {\sc Li}(1993)}]{FuLi:1993}
{\sc Fu, Y.}, and {\sc W.~Li}, 1993 Statistical tests of neutrality of
  mutations.
\newblock Genetics {\bf 133}: 693--709.

\bibitem[{{\sc Garrigan} and {\sc Hammer}(2006)}]{GarriganHammer:2006}
{\sc Garrigan, D.}, and {\sc M.~F. Hammer}, 2006 Reconstructing human origins
  in the genomic era.
\newblock Nat. Rev. Genet. {\bf 7}: 669--680.

\bibitem[{{\sc Griffiths} and {\sc Tavar\'e}(1994)}]{GriTav:94}
{\sc Griffiths, R.}, and {\sc S.~Tavar\'e}, 1994 Sampling theory for neutral
  alleles in a varying environment.
\newblock Phil. Trans. Roy. Soc. Lon. B {\bf 344}: 403--410.

\bibitem[{{\sc Jagers} and {\sc Sagitov}(2004)}]{JagersSagitov:2004}
{\sc Jagers, P.}, and {\sc S.~Sagitov}, 2004 Convergence to the coalescent in
  populations of substantially varying size.
\newblock Journal of Applied Probability {\bf 41}: 368--378.

\bibitem[{{\sc Johannesson}(2009)}]{Johannesson:2009}
{\sc Johannesson, K.}, 2009 private communication .

\bibitem[{{\sc Kaj} and {\sc Krone}(2003)}]{Kaj:2003}
{\sc Kaj, I.}, and {\sc S.~Krone}, 2003 The coalescent process in a population
  with stochastically varying size.
\newblock J. Appl. Prob. {\bf 40}: 33--48.

\bibitem[{{\sc Kimmel} and {\sc Chakraborty}(1996)}]{kimmel1996measures}
{\sc Kimmel, M.}, and {\sc R.~Chakraborty}, 1996 Measures of variation at {DNA}
  repeat loci under a general stepwise mutation model.
\newblock Theoretical Population Biology {\bf 50}: 345--367.

\bibitem[{{\sc Kingman}(1982)}]{Kingman:1982}
{\sc Kingman, J.}, 1982 The coalescent.
\newblock Stoch. Proc. Appl. {\bf 13}: 235--248.

\bibitem[{{\sc Moran}(1958)}]{Moran:1958}
{\sc Moran, P.}, 1958 Random processes in genetics.
\newblock Proc. Cambridge Philos. Soc. {\bf 54}: 60--71.

\bibitem[{{\sc Nordborg} and {\sc Krone}(2003)}]{Nordborg:2003}
{\sc Nordborg, M.}, and {\sc S.~Krone}, 2003 {\em Modern Developments in
  Population Genetics: The Legacy of Gustave Mal\'e{}cot\/}.
\newblock Oxford University Press, Oxford, 194--232.

\bibitem[{{\sc Ohta} and {\sc Kimura}(1973)}]{Ota:1973}
{\sc Ohta, T.}, and {\sc M.~Kimura}, 1973 A model of mutation appropriate to
  estimate the number of electrophoretically detectable alleles in a finite
  population.
\newblock Genet Res {\bf 22}: 201--204.

\bibitem[{{\sc Sagitov} {\em et~al.\/}(2010){\sc Sagitov}, {\sc Rafajlovic},
  {\sc Mehlig} and {\sc Eriksson}}]{Serik:2010}
{\sc Sagitov, S.}, {\sc M.~Rafajlovic}, {\sc B.~Mehlig}, and {\sc A.~Eriksson},
  2010 External branch lengths of genealogies in expanding and in declining
  populations.
\newblock unpublished .

\bibitem[{{\sc Sj{\"o}din} {\em et~al.\/}(2005){\sc Sj{\"o}din}, {\sc Kaj},
  {\sc Krone}, {\sc Lascoux} and {\sc Nordborg}}]{Sjodin:2005}
{\sc Sj{\"o}din, P.}, {\sc I.~Kaj}, {\sc S.~Krone}, {\sc M.~Lascoux}, and {\sc
  M.~Nordborg}, 2005 On the meaning and existence of an effective population
  size.
\newblock Genetics {\bf 169}: 1061--1070.

\bibitem[{{\sc Slatkin}(1996)}]{Slatkin}
{\sc Slatkin, M.}, 1996 Gene genealogies within mutant allelic classes.
\newblock Genetics {\bf 143}: 579--587.

\bibitem[{{\sc Tajima}(1989)}]{taj89:sta}
{\sc Tajima, F.}, 1989 Statistical method for testing the neutral mutation
  hypothesis by {DNA} polymorphism.
\newblock Genetics {\bf 123}: 585--595.

\bibitem[{{\sc Tavar\'e}(1984)}]{Tav:84}
{\sc Tavar\'e, S.}, 1984 Evolution of coalescence times, genetic diversity and
  structure during colonization.
\newblock Theor. Pop. Biol. {\bf 26}: 119--164.

\bibitem[{{\sc Tavar\'e}(2004)}]{Tav:2004}
{\sc Tavar\'e, S.}, 2004 {\em Ancestral Inference in Population Genetics\/}.
\newblock Springer, Berlin, 1--188.

\bibitem[{{\sc Wakeley} and {\sc Sargsyan}(2009)}]{Wakeley:2009}
{\sc Wakeley, J.}, and {\sc O.~Sargsyan}, 2009 Extensions of the coalescent
  effective population size.
\newblock Genetics {\bf 181}: 341--345.

\bibitem[{{\sc Watterson}(1975)}]{Watterson:1975}
{\sc Watterson, G.~A.}, 1975 On the number of segregation sites in genetical
  models without recombination.
\newblock Theor. Pop. Biol. {\bf 7}: 256--276.

\bibitem[{{\sc Wright}(1931)}]{Wright:1931}
{\sc Wright, S.}, 1931 Evolution in mendelian populations.
\newblock Genetics {\bf 16}: 97--159.

\bibitem[{{\sc Zeng} {\em et~al.\/}(2006){\sc Zeng}, {\sc Fu}, {\sc Shi} and
  {\sc Wu}}]{ZengFuShiWu:2006}
{\sc Zeng, K.}, {\sc Y.~Fu}, {\sc S.~Shi}, and {\sc C.~Wu}, 2006 Statistical
  tests for detecting positive selection by utilizing high-frequency variants.
\newblock Genetics {\bf 174}: 1431--1439.

\bibitem[{{\sc Zivkovic} and {\sc Wiehe}(2008)}]{Ziv:2008}
{\sc Zivkovic, D.}, and {\sc T.~Wiehe}, 2008 Second-order moments of
  segregating sites under variable population size.
\newblock Genetics {\bf 180}: 341--357.

\end{thebibliography}

\newpage
\appendix

\section{Alternative calculation of $F_n(q)$ and $\langle T_n^k\rangle$}
\label{sec:A}
In this appendix we demonstrate and alternative way of calculating
function $F_n(q)$ and the moments $\langle T_n^k\rangle$ within 
the coalescent approximation. 
\begin{figure}
\centerline{\includegraphics[angle=0,width=9cm,clip]{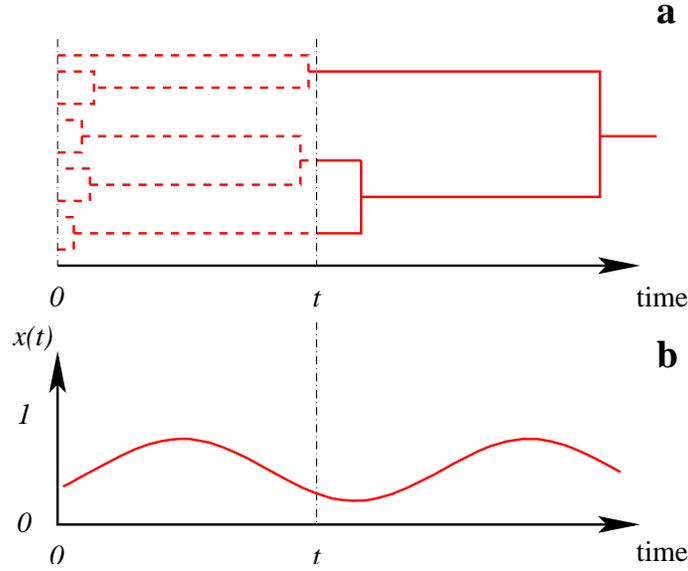}}
\caption{\label{fig:theta2} {\bf a} Illustrates the definition of 
$F_n(q,t)$  which for $q=\theta/2$ is the probability that $n$
sequences sampled at time $t$ are identical.
The corresponding branches in the tree
are drawn as solid lines, and $n=3$ in the figure.
{\bf b} Shows schematically how the population size $x(t)$ depends upon time.}
\end{figure}
Given a realisation of the curve $x(t)$, the function $F_n(q)$ can
be calculated as follows. 

Consider the function $F_n(q,t)$ that is, for $q=\theta/2$, the probability
that $n$ sequences sampled at time $t$ are identical. The time argument
describes how $F_n$ depends upon the time at which the population
is sampled, given a smooth population-size curve $x(t)$.
The definition of $F_n(q,t)$  is illustrated
for the case $n=9$ in Fig.~\ref{fig:theta2}. 
In Secs. \ref{sec:rec} and \ref{sec:num}, the populations
were sampled at $t=0$, which corresponds to the choice $F_n(q) = F_n(q,0)$.
The more general quantity $F_n(q,t)$ allows to determine how the fluctuations
of sample genealogies depend upon the time of sampling (in a population
of constant size, $F_n$ is independent of $t$).

To obtain a recursion 
for $F_n(q,t)$ in a population of fluctuating size, take $q = \theta/2$ and consider
a small time interval $\delta t$. A change in $F_n$ during this time interval is due to either a mutation in one of the ancestral lines, or to two ancestral lines having a common ancestor. Thus, to first order in $\delta t$
\bal{\label{eq:Fn_der_1}
        F_n(q,t) = \left[1 - nq\delta t 
-  \frac{n (n-1)}{2 x(t)}\delta t\right] F_n(q,t + \delta t) 
+  \frac{n (n-1)}{2 x(t)}\delta t F_{n-1}(q,t+\delta t).
}
Taking the limit $\delta t \ra 0$, 
we obtain:
\bal{\label{eq:Fn_der_2}
        \frac{\partial}{\partial t}F_n(q,t) = 
\left[nq + \frac{n (n-1)}{2 x(t)}\right] F_n(q,t) - \frac{n (n-1)}{2 x(t)} F_{n-1}(q,t).
}
The recursion is terminated by $F_1(q,t) \equiv 1$ for all values of $t$. 
In a population of constant size $x=1$, $F_n(q,t)$ 
does not depend upon $t$ and
the result (\ref{eq:Fn}) 
is immediately recovered from (\ref{eq:Fn_der_2}).
To find the general solution, Eq.~(\ref{eq:Fn_der_2}) 
is rewritten as follows:
\begin{equation}
F_n(q,t) = b_n\int_t^\infty\!\!\frac{{\rm d}s }{x(s)}
{\rm e}^{nq(t-s)+b_n [\Lambda(t)-\Lambda(s)]} F_{n-1}(q,s)\,.
\end{equation}
It is convenient to consider the function $G_n(q,t) = {\rm e}^{-nqt-b_n\Lambda(t)} F_n(q,t)$. It obeys the recursion
\begin{equation}
G_n(q,t) = b_n\int_t^\infty\!\!\frac{{\rm d} s}{x(s)}  
{\rm e}^{-qs-(n-1)\Lambda(s)} G_{n-1}(q,s) \equiv (\LL_n \phi G_{n-1})(t)\,.
\end{equation}
Here $\LL_k$ is the operator 
\begin{equation}
\label{eq:defL}
(\LL_k f)(t) =
b_k 
\int_t^\infty\!\!\frac{{\rm d}s}{x(s)} {\rm e}^{-(k-1)\Lambda(s)} f(s)\,,
\end{equation}
and $\phi(s) = \exp(-q s)$.
In terms of this operator, the recursion is simply solved by
repeated action of $\LL_k$ on the function $\phi$:
\begin{equation}
\label{eq:Gn_nest_a}
        G_n(q,t) = (\LL_n \phi \LL_{n-1} \cdots \phi\LL_2 \phi^2)(t)\,,
\end{equation}
where we have used the fact that $G_1(q,s) = \exp(-qs)$.
Noting that $G_n(q,0) = F_n(q,0)\equiv F_n(q)$ we obtain
the desired expression (\ref{eq:r2}) for $F_n(q)$:
\begin{equation}
\label{eq:Fn_nest}
 F_n(q) = (\LL_n \phi \LL_{n-1} \cdots \phi\LL_2 \phi^2)(0)\,.
\end{equation}

The moments $\langle T_n^k\rangle$ can be calculated in a similar fashion.
Consider the function $\langle T_n^k(t)\rangle$ describing the moments
of the total length of all solid branches shown in Fig. \ref{fig:theta2}.
Then $\langle T_n^k\rangle = \langle T_n^k(0)\rangle$, and $\langle T_n^k(t)\rangle$
obeys the recursion:
\bal{\label{eq:mom_1}
	\frac{\partial}{\partial t} \expt{T^k_n(t)} =   - k n \expt{T^{k-1}_n(t)} + 
	      \frac{b_n}{x(t)}\Big[ \expt{T^k_n(t)} - \expt{T^k_{n-1}(t)} \Big]\,.
}
We introduce the function $h^{(k)}_n(t) = e^{-b_n \Lambda(t)}\expt{T^k_n(t)}$.
With the initial conditions
$h^{(0)}_n(t) = e^{-b_n\Lambda(t)}$ for $n \ge 2$ and $h^{(m)}_1(t) = 0$ for $m \ge 1$, 
the recursion (\ref{eq:mom_1}) is simply:
\bal{\label{eq:mom_5}
	h^{(k)}_n(t) = k \sum_{m=2}^{n} m  (\LL_n \LL_{n-1} \cdots \LL_{m+1} \II  h^{(k-1)}_m)(t)
}
where $\II$ is the integral operator $\int_t^\infty {\rm d}t'\,$. 

Now we show that the action of the chain $\LL_n \LL_{n-1} \cdots \LL_{m+1}$ of operators
on an arbitrary function $f$ can be represented in terms of a single integral.
To show this, it is convenient 
to make a change of variables to $z=\Lambda(t)$:
\begin{equation}
\label{eq:kernel}
(\LL_kf)(z) = b_k \int_z^\infty {\rm d}y\, {\rm e}^{-(k-1) y} f(y)\,.
\end{equation}
The task is to
seek a kernel $K_{nm}(z,z')$
such that for any function $f$
\bal{\label{eq:prop_der_1}
(\LL_{n}\LL_{n-1}\cdots\LL_{m}{f})(z) = 
\int_z^\infty {\rm d}z' K_{nm}(z,z') {f}(z')\,.
}
The kernel must satisfy 
\bal{\label{eq:prop_der_2b}
        K_{nm}(z,z') = b_n \int_z^{z'}\!\! {\rm d}y\,  e^{-(n-1)y} K_{n-1,m}(y,z') 
}
Together with the initial condition $K_{mm}(z,z') = b_m \exp[-(m-1)z']\, H(z'-z)$, this recursion
allows to compute the kernel in closed form. This can for example be achieved
by considering the Laplace transform of  (\ref{eq:kernel}). We find:
\begin{eqnarray}
\label{eq:prop_der_7}
        K_{nm}(z,z')&\!=\!&\sum _{j=m}^n  k_{nmj}
        {\rm e}^{-(b_n - b_j)z} \, {\rm e}^{(b_{m-1} - b_j)z'}\\
        k_{nmj}&\!=\!&(-1)^{j-m} \frac{2j\!-\!1}{2}
        \frac{\Gamma(m+j-1)}{\Gamma(m)\Gamma(m-1)\Gamma(j-m+1)}
        \frac{\Gamma(n)\Gamma(n+1)}{\Gamma(n+j)\Gamma(n-j+1)}\,.
\nonumber
\end{eqnarray}
This kernel can be used to evaluate (\ref{eq:mom_5}).
For any function $g(t)$ we have that
\bal{\label{eq:mom_7}
	(\LL_{n}\cdots \LL_{m+1} \II g )(t)
	&= \int_t^\infty {\rm d}t'\, A_{n m+1}\big(\Lambda(t),\Lambda(t')\big) g(t')
}
with
\bal{
\label{eq:m_7}
	A_{nm+1}(z,z') &= \int_z^{z'} {\rm d}z'' K_{nm+1}(z,z'') 
} 
(and $A_{nn+1}(z,z') = 1$). Inserting
this result into (\ref{eq:mom_5}) yields      
\begin{equation}
h_n^{(k)}(t) = k \sum_{m=2}^n m \int_t^\infty{\rm d}t' A_{nm+1}\big(
\Lambda(t'),\Lambda(t)\big)\, h_m^{(k-1)}(t')\,.
\end{equation}
Identifying $A_{nm+1}(z,z') =  \exp(-b_n z) g_{nm}(z'-z) \exp(b_m z')$
we find
\begin{equation}
\label{eq:mom_9}
\langle T_n^{k}(t)\rangle = 
k \sum_{m=2}^n m \int_t^\infty{\rm d}t' f_{nm}(t,t') 
\langle T_m^{k-1}(t')\rangle \,.
\end{equation}
This recursion yields Eq. (\ref{eq:corr}).

\newpage
\section{Coefficients $d_{n;j_1,j_2}$ for $n=2,\ldots,10$}
\label{sec:B}
In Tab. \ref{tab:1} we give the coefficients $d_{n;j_1,j_2}$ determining
the second moment $\langle T_n^2\rangle$ according to Eq.~(\ref{eq:corr2})
for $n=2,\ldots,10$.
Note that the coefficient for $n=2$ is consistent with Eq.~(\ref{eq:T2k}).
\begin{table}[b]
\includegraphics[angle=0,width=7.cm,clip]{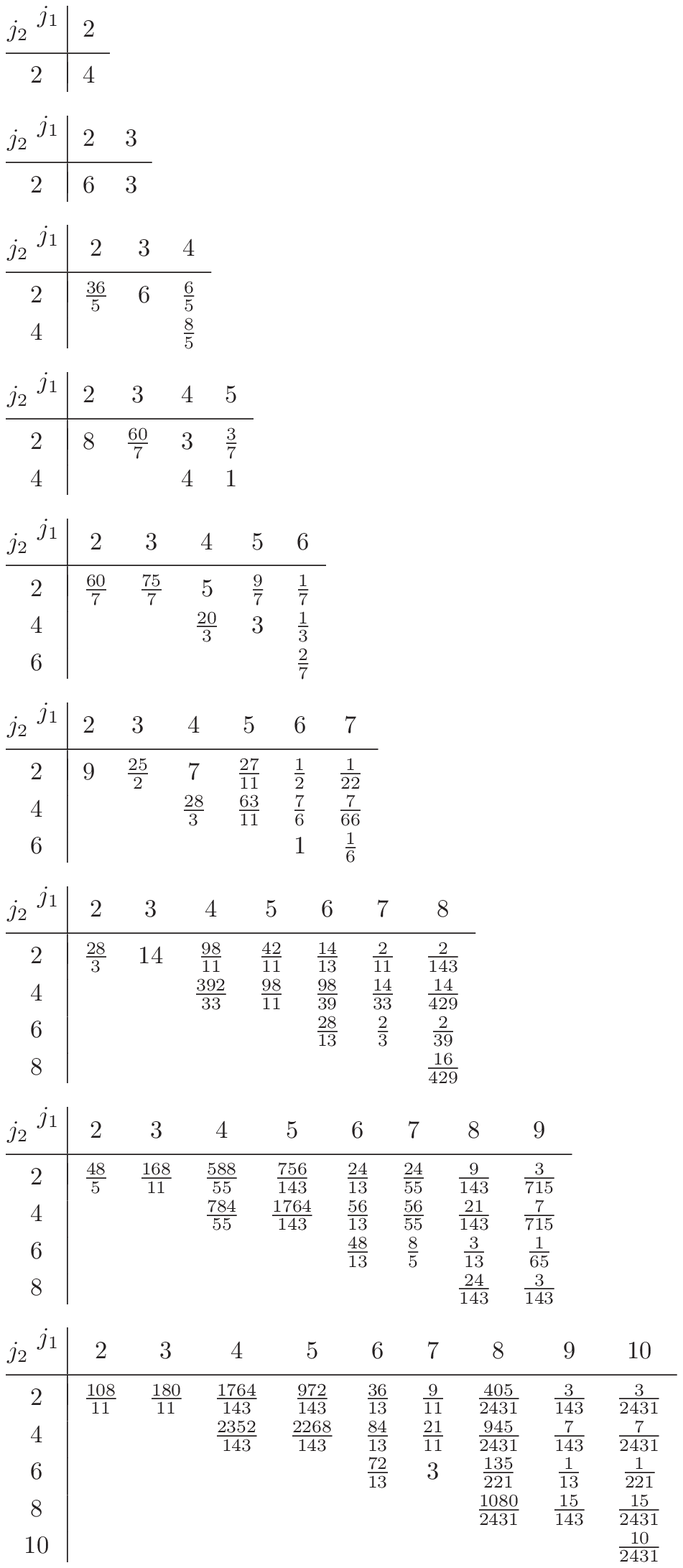}
\caption{\label{tab:1}
Shows coefficients $d_{n;j_1,j_2}$ occurring in Eq.~(\ref{eq:corr2}) for $n=2,\ldots,10$.
Coefficients for odd values of $j_2$ vanish.}
\end{table}

\end{document}